\documentclass[12pt]{revtex4}
\usepackage{dcolumn}
\usepackage{graphicx}
\usepackage{amsmath}
\usepackage{amsfonts}
\usepackage{amssymb}
\usepackage{psfrag}
\usepackage{wrapfig}
\usepackage{subfigure}
\usepackage{makeidx}
\usepackage{bm}
\usepackage{epsf}
\usepackage{hyperref}

\begin{document}

\title{Thermodynamics of Gauss-Bonnet-Born-Infeld black holes in AdS space}

\author{De-Cheng Zou$^{1}$, Zhan-Ying Yang$^{1}$\footnote{Email:zyyang@nwu.edu.cn},
Rui-Hong Yue$^{2}$\footnote{ Email:yueruihong@nbu.edu.cn}
and Peng Li$^3$}
\affiliation{ $^{1}$Department of Physics, Northwest University, Xi'an, 710069, China\\
$^{2}$Faculty of Science, Ningbo University, Ningbo 315211, China\\
$^{3}$Institute of Modern Physics, Northwest University, Xi'an, 710069, China
}

\date{\today}

\begin{abstract}
\indent

We construct solutions of a model which includes the Gauss-Bonnet and Born-Infeld terms
for various horizon topologies $(k=0, \pm 1)$, and then the mass, temperature, entropy
and heat capacity of black holes are computed. For the sake of simplicity, we perform
the stability analysis of five dimensional topological black holes in AdS space.
\end{abstract}


\maketitle

\section{Introduction}
\label{intro}
A renewed interest in the Lovelock gravity and Born-Infeld electrodynamics has appeared
because they emerge in the low energy limit of string
theory \cite{Fradkin:1985qd, Bergshoeff:1987at, Metsaev:1987qp}.
The effect of string theory on the left hand side of field equations of gravity is
usually investigated by means of a low energy effective action which describes gravity
at the classical level \cite{Lust:1989tj, Myers:1987yn, Alekseev:1997wy, Maeda:2009uy}.
In addition to Einstein-Hilbert action,
this effective action also involves higher derivative curvature terms.
In the AdS/CFT correspondence, these higher derivative curvature
terms correspond to the correction terms of large N expansion in the CFT
side \cite{Dehghani:2009zzb}. In general, the higher
powers of curvature could give rise to a fourth or even higher order differential
equation for the metric, and it would introduce ghosts and violate
unitarity, therefore, the higher derivative terms may be a source of inconsistencies.
However, the so-called Lovelock gravity is quite special \cite{Lovelock:1971yv}.
Its Lagrangian is the sum of dimensionally extended Euler densities
\begin{eqnarray}
{\cal L}=\sum^{m}_{n=0}\alpha_n{\cal L}_n,\label{eq:1a}
\end{eqnarray}
where $\alpha_n$, arbitrary constants, are the Lovelock coefficients and ${\cal L}_n$
is the Euler density of a $2k$-dimensional manifold
\begin{eqnarray}
{\cal L}_n=\frac{1}{2^n}\delta^{a_1b_1\cdots a_nb_n}_{c_1b_1\cdots c_nd_n}
R^{c_1d_1}_{~~~~a_1b_1}\cdots R^{c_nd_n}_{~~~~a_nb_n}.\label{eq:2a}
\end{eqnarray}
Here the generalized delta function $\delta^{a_1\cdots b_n}_{c_1\cdots d_n}$ is totally
antisymmetric in both sets of indices and $R^{cd}_{~~ab}$ is the Riemann tensor. Though
the Lagrangian of Lovelock gravity consists of some higher derivative curvature terms,
its field equations of motion contain the most symmetric conserved tensor with no more
than two derivative of the metric. They have also been shown to be ghost-free when
expanding about flat space, evading any problem with unitarity \cite{Zwiebach:1985uq, Wheeler:1985qd}.
In this paper, we indulge ourselves to with the first three terms of the Lovelock
gravity, corresponding to the cosmological term, Einstein and Gauss-Bonnet terms, respectively.
It has been argued that the Gauss-Bonnet term appears as the leading correction
to the low energy effective action of the string theory and its Lagrangian is given by
\begin{eqnarray}
{\cal L}_{GB}=R_{\gamma\delta\lambda\sigma}R^{\gamma\delta\lambda\sigma}
-4R_{\gamma\delta}R^{\gamma\delta}+R^2.\label{eq:3a}
\end{eqnarray}
In Gauss-Bonnet gravity, static and spherically symmetric black hole solutions were firstly
presented in \cite{Boulware:1985wk}, and of the charged in \cite{Wiltshire:1985us}. The
thermodynamics of these solutions have been investigated in \cite{Myers:1988ze}, of
solutions with nontrivial topology in \cite{Cai:2001dz}
and the charged black hole solutions in \cite{Cvetic:2001bk}.

Besides the curvature terms, one would also expect higher derivative gauge field
contributions to the action. This is done by explicitly constructing black holes
solutions coupled to a Born-Infeld gauge field in the presence of a
cosmological constant. The Born-Infeld electrodynamics is the nonlinear generalization
of the Reissner-Nordstr\"{o}m black hole (RNAdS) and is characterized by charged $Q$,
mass $M$ and the nonlinear parameter $\beta$. Its Lagrangian ${\cal L(F)}$ is given by
\begin{eqnarray}
{\cal L(F)}=4\beta^2(1-\sqrt{1+\frac{F^{\mu\nu}F_{\mu\nu}}{2\beta^2}}),\label{eq:4a}
\end{eqnarray}
where the constant $\beta$ is the Born-Infeld parameter,
$F_{\mu\nu}=\partial_{\mu}A_{\nu}-\partial_{\nu}A_{\mu}$ is electromagnetic tensor
field and $A_{\mu}$ is the vector potential. The Born-Infeld theory was originally
introduced to get a classical theory of charged particles with finite
self-energy \cite{Born:1934gh}. Hoffmann \cite{Hoffmann:1935ty} was the first
one in relating general relativity and the Born-Infeld electromagnetic
field. He obtained a solution of the Einstein equations for a point-like Born-Infeld
charge, while is devoid of the divergence of the metric at the origin that
characterizes the Reissner-Nordstr\"{o}m solution \cite{Aiello:2004rz}. But, a
conical singularity remained there, as it was later objected by Einstein and Rosen.
The Born-Infeld black hole with zero cosological constant was obtained by Garcia $et$
$al$ \cite{Garica:1984}. Later, The spherically symmetric Einstein-Born-Infeld black
hole solutions with cosmological constant were studied in \cite{Cai:2004eh},
Born-Infeld-dilaton models in \cite{Sheykhi:2008rt}. Note that the work of the
BIAdS in the grand canonical ensemble was carried out in \cite{Myung:2008eb}.
The Euclidean action for the grand canonical ensemble was computed with the appropriate
boundary terms. The thermodynamical quantities such as the Gibbs free energy, entropy
and specific heat of the black holes are derived from it. For the Lovelock gravity,
the asymptotically flat Born-Infeld black hole solutions in Gauss-Bonnet gravity were
found in \cite{Wiltshire:1988uq}, five dimensional AdS black hole solutions in \cite{Aiello:2004rz}.
The Born-Infeld black hole solutions in third order Lovelock gravity have been obtained
in \cite{Dehghani:2008qr}. The rotating Born-Infeld black hole solutions have been analyzed
by Dehghani $et$ $al$ in general relativity \cite{Dehghani:2006zi}, Gauss-Bonnet
gravity \cite{Dehghani:2006ke} and third order Lovelock gravity \cite{Dehghani:2008qr},
respectively. In this paper, we are concerned with the Born-Infeld-anti-de Sitter black hole (BIAdS)
in Gauss-Bonnet gravity and discuss the thermodynamic quantities of five dimensional
black holes in the canonical ensemble.

The structure of this paper is as follows. In section \ref{22s}, we present $n+1$
dimensional BIAdS black hole solutions in Gauss-Bonnet gravity.
Then, the thermodynamics of these black holes will be discussed for a fixed-charge.
In order to simplify the analysis of the thermodynamic properties, we study
Gauss-Bonnet-Born-Infeld black holes in five dimensional spacetimes and investigate the
stability of black holes by computing heat capacity of black holes in section \ref{33s}.
Section \ref{44s} devotes to concluding remarks.

\section{Gauss-Bonnet-Born-Infeld black holes in AdS space}
\label{22s}
The action of Gauss-Bonnet gravity in the presence of nonlinear Born-Infeld electromagnetic
field can be written as
\begin{eqnarray}
{\mathbf I}=\frac{1}{16\pi}\int d^{n+1}x \sqrt{-g}[-2\Lambda+R+\alpha {\cal L}_{GB}
+{\cal L(F)}],\label{eq:5a}
\end{eqnarray}
where $\Lambda=-\frac{n(n-1)}{2l^2}$ is a negative cosmological constant, $\alpha$ is the
Gauss-Bonnet coefficient with dimension $(length)^2$ and is positive in the heterotic
string theory. In the limit $\beta\rightarrow \infty$, ${\cal L(F)}$ Eq.~(\ref{eq:4a})
reduces to the standard Maxwell form
\begin{eqnarray}
{\cal L(F)}=-F^{\mu\nu}F_{\mu\nu}+\mathcal {O}(F^4).\label{eq:6a}
\end{eqnarray}
By varying the action Eq.~(\ref{eq:5a}) with regard to the gauge field $A_\mu$ and
the metric $g_{\mu\nu}$, these corresponding equations of motion are expressed as
\begin{eqnarray}
\partial_{\mu}\left(\frac{\sqrt{-g}F^{\mu\nu}}{\sqrt{1
+\frac{F^2}{2\beta^2}}}\right)=0\label{eq:7a}
\end{eqnarray}
and
\begin{eqnarray}
R_{\mu\nu}-\frac{1}{2}Rg_{\mu\nu}&=&\frac{n(n-1)}{2l^2}g_{\mu\nu}
+\alpha[\frac{1}{2}g_{\mu\nu}(R_{\gamma\delta\lambda\sigma}R^{\gamma\delta\lambda\sigma}
-4R_{\gamma\delta}R^{\gamma\delta}+R^2)\nonumber\\
&-&2RR_{\mu\nu}+4R_{\mu\gamma}R^{\gamma}_{~\nu}+4R_{\gamma\delta}R^{\gamma~\delta}_{~\mu~\nu}
-2R_{\mu\gamma\delta\lambda}R_{\nu}^{~\gamma\delta\lambda}]\nonumber\\
&+&\frac{2F_{\mu\lambda}F_{\nu}^{~\lambda}}{\sqrt{1+\frac{F_{\mu\lambda}F_{\nu}^{~\lambda}}{2\beta^2}}}
+\frac{1}{2}g_{\mu\nu}{\cal L(F)}.\label{eq:8a}
\end{eqnarray}

We assume the metric being of the following form
\begin{eqnarray}
ds^2=-f(r)dt^2+\frac{1}{f(r)}dr^2+r^2h_{ij}dx^idx^j,\label{eq:9a}
\end{eqnarray}
where the coordinates are labelled as $x^{\mu}=(t, r, x^i)$, $(i=1,\cdots , (n-1))$. The metric function
$h_{ij}$ is a function of the coordinates $x^i$ which span an $(n-1)$-dimensional hypersurface with
constant scalar curvature $(n-1)(n-2)$k. The constant $k$ characterizes the geometric property of
hypersurface and takes values $k=0$ (flat), $k=-1$ (negative curvature) and $k=1$ (positive curvature).
$f(r)$ is an unknown function of $r$ which will determined later.

In the static and symmetric background Eq.~(\ref{eq:9a}), a class of solutions of Eq.~(\ref{eq:7a})
can be written down where all the components of $F^{\mu\nu}$ are zero except $F^{rt}$ \cite{Cai:2004eh}
\begin{eqnarray}
F^{rt}=\frac{\sqrt{(n-1)(n-2)}\beta q}{\sqrt{2\beta^2r^{2n-2}+(n-1)(n-2)q^2}},\label{eq:10a}
\end{eqnarray}
where $q$ is an integration constant relating to the electric charged of the solution. This can be verify
from the behavior of $F^{rt}$ in the limit $\beta \rightarrow \infty$ as $F^{rt} \sim \frac{q}{r^{n-1}}$.
Via $Q=\frac{1}{4\pi}\int *F d\Omega$, the charge of the black hole can be found by compute the
flux of the electric field at infinity
\begin{eqnarray}
Q=\frac{q\Sigma_k }{4\pi}\sqrt{\frac{(n-1)(n-2)}{2}},\label{eq:11a}
\end{eqnarray}
where $\Sigma_k$ represents the volume of constant curvature hypersurface described by $h_{ij}dx^idx^j$.

Substituting the metric Eq.~(\ref{eq:9a}) and Eq.~(\ref{eq:10a}) into Eq.~(\ref{eq:8a}), one may use
any components of Eq.~(\ref{eq:8a}) to find the function $f(r)$. The simplest equation is
the $rr$-component of these equations which can be written as
\begin{eqnarray}
\frac{n(n-1)}{l^2}&-&4\beta^2(\sqrt{1+\eta}-1)=f(r)'[2\alpha(n-2)(n-3)(k-f(r))/r^3\nonumber\\
&+&1/r](n-1)-[\alpha(n-3)(n-4)(k-f(r))/r^4\nonumber\\
&+&1/r^2](n-1)(n-2)(k-f(r)),\label{eq:12a}
\end{eqnarray}
where prime denotes the derivative with regard to $r$ and the function $\eta$ is
$\frac{(n-1)(n-2)}{2\beta^2r^{2n-2}}q^2$.
Then, the solution of Eq.~(\ref{eq:12a}) is given by
\begin{eqnarray}
f(r)=k+\frac{r^2}{2\tilde{\alpha}}(1\pm\sqrt{g(r)}),\label{eq:13a}
\end{eqnarray}
where the function $g(r)$ is demonstrated by using a hypergeometric function
\begin{eqnarray}
g(r)&=&1-\frac{4\tilde{\alpha}}{l^2}+\frac{4\tilde{\alpha} m}{r^n}-\frac{16\tilde{\alpha}\beta^2}{n(n-1)}\nonumber\\
&+&\frac{8\sqrt{2}\tilde{\alpha}\beta}{n(n-1)r^{n-1}}\sqrt{2\beta^2r^{2n-2}+(n-1)(n-2)q^2}\nonumber\\
&-&\frac{8(n-1)\tilde{\alpha}q^2}{nr^{2n-2}}\times _2F_1[\frac{n-2}{2n-2},\frac{1}{2},\frac{3n-4}{2n-2},
-\frac{(n-1)(n-2)q^2}{2\beta^2 r^{2n-2}}].\nonumber
\end{eqnarray}
Here $\tilde{\alpha}=(n-2)(n-3)\alpha$ and $m$ is an integration constant which is related to
mass of the solution. In our convention, the ADM mass $M$ is $\frac{(n-1)\Sigma_km}{16\pi}$.
Obviously, the Gauss-Bonnet-BIAdS black hole is characterized by charged $Q$, mass $M$, nonlinear parameter $\beta$,
parameter $l$ and the coefficient $\tilde{\alpha}$ of Gauss-Bonnet term. Note that the function $g(r)$
in Eq.~(\ref{eq:13a}) is negative for very small $r$. In order to maintain
the non-negativity for $g(r)$, we set that $r$ is larger than a minimum $r (r_{min})$, where $r_{min}$ is the root of $g(r)$.

The hyper-surface $r=r_{min}$ is essentially a curvature singularity and the given solution will be a black hole
solution if this singular hyper-surface is surrounded by the event horizon (having radius $r_+$
such that $f(r_+)=0$), otherwise, the solution describes a naked singularity. In addition,
here we only focus on the positive coefficient $\tilde{\alpha}$. When $k=1$ and $1/l^2=0$,
the solution is asymptotically flat solution if one take "-" sign. But it is asymptotically anti-de Sitter
solution with a negative gravitational mass for the "+" sign, indicating the instability.
Hence, we will only consider the branch with "-" branch. It can be checked that for $n=3$, it reduces to
the solution of \cite{Aiello:2004rz}.

Based on Eq.~(\ref{eq:10a}), the electric field $F_{rt}$ is obtained
\begin{eqnarray}
F_{rt}=-\frac{\sqrt{(n-1)(n-2)}\beta q}{\sqrt{2\beta^2r^{2n-2}+(n-1)(n-2)q^2}}.\label{eq:14a}
\end{eqnarray}
Then, according to $F_{rt}=\partial_{r}A_{t}-\partial_{t}A_{r}$, the electric gauge potential
is given by
\begin{eqnarray}
A_t=\sqrt{\frac{n-1}{2n-4}}\frac{q}{r^{n-2}}F(\eta)-\Phi,\label{eq:15a}
\end{eqnarray}
where $\Phi$ is a constant and function $F(\eta)$ is
$F(\eta)=_2F_1[\frac{n-2}{2n-2},\frac{1}{2},\frac{3n-4}{2n-2},-\frac{(n-1)(n-2)q^2}{2\beta^2 r^{2n-2}}]$.
In order to fix the gauge potential at the horizon to be zero,
the quantity $\Phi$ is chosen as
\begin{eqnarray}
\Phi=\sqrt{\frac{n-1}{2n-4}}\frac{q}{r_+^{n-2}}F(\eta_+).\label{eq:16a}
\end{eqnarray}
Clearly, the quantities $A_t$ and $\Phi$ in Gauss-Bonnet gravity is identical with
corresponding ones \cite{Cai:2004eh} in general relativity.

In the limit $\tilde{\alpha}\rightarrow 0$, the black hole solution Eq.~(\ref{eq:13a}) tends to
\begin{eqnarray}
f(r)&=&k+\frac{r^2}{l^2}-\frac{m}{r^{n-2}}+\frac{4\beta^2r^2}{n(n-1)}
-\frac{2\sqrt{2}\beta}{n(n-1)r^{n-3}}\nonumber\\
&\times&\sqrt{2\beta^2r^{2n-2}+(n-1)(n-2)q^2}+\frac{2(n-1)q^2}{nr^{2n-4}}F(\eta).\label{eq:17a}
\end{eqnarray}
one can find that it corresponds to the Einstein-Born-Infeld black hole solution in AdS
space \cite{Cai:2004eh}. While, in the limit $q \rightarrow 0$, the solution Eq.~(\ref{eq:13a})
reduces to uncharged Gauss-Bonnet solution
\begin{eqnarray}
f(r)=k+\frac{r^2}{2\tilde{\alpha}}(1-\sqrt{1-\frac{4\tilde{\alpha}}{l^2}
+\frac{4\tilde{\alpha}m}{r^n}}).\label{eq:18a}
\end{eqnarray}
On the other hand, if taking $\beta \rightarrow \infty$ in Eq.~(\ref{eq:13a}),
the solution $f(r)$ reduces to the Lovelock-Maxwell solution found in \cite{Wiltshire:1985us}
\begin{eqnarray}
f(r)=1+\frac{r^2}{2\tilde{\alpha}}(1-\sqrt{1-\frac{4\tilde{\alpha}}{l^2}
+\frac{4\tilde{\alpha}m}{r^n}-\frac{4\tilde{\alpha}q^2}{r^{2n-2}}}).\nonumber
\end{eqnarray}

In the rest of the section, we discuss the thermodynamic properties of black holes.
In terms of horizon radius $r_+$, the ADM mass $M$ is obtained
\begin{eqnarray}
M&=&\frac{(n-1)\Sigma_kr_+^{n-2}}{16\pi}\left\{k+\frac{\tilde{\alpha}k^2}{r_+^2}+\frac{r_+^2}{l^2}
+\frac{4\beta^2r_+^2}{n(n-1)}-\frac{2\sqrt{2}\beta}{n(n-1)r_+^{n-3}}\right.\nonumber\\
&\times&\left.\sqrt{2\beta^2r_+^{2n-2}+(n-1)(n-2)q^2}+\frac{(n-1)}{n}\frac{2q^2}{r_+^{2n-4}}
F(\eta_+)\right\}.\label{eq:19a}
\end{eqnarray}
The Hawking temperature of black holes can be obtained by
requirement of the absence of conical singularity at the event horizon in the Euclidean
section of the black hole solution. It can be written
as $T_H=\frac{f'(r_+)}{4\pi}$
\begin{eqnarray}
T_H&=&\frac{1}{4\pi r_+(r_+^2+2k\tilde{\alpha})}\left\{\frac{nr_+^4}{l^2}+(n-2)kr_+^2
+(n-4)k^2\tilde{\alpha}+\frac{4\beta^2r_+^4}{n-1}\right.\nonumber\\
&-&\left.\frac{2\sqrt{2}\beta}{(n-1)r_+^{n-5}}
\times\sqrt{2\beta^2r_+^{2n-2}+(n-1)(n-2)q^2}\right.\}.\label{eq:20a}
\end{eqnarray}
We note that in the limit $\beta \rightarrow \infty$ and $q \neq 0$, $T_H$ reduces to the charged case
\begin{eqnarray}
\tilde{T}_H=\frac{nr_+^4+(n-2)kl^2r_+^2+(n-4)k^2\tilde{\alpha}l^2
-(n-2)q^2l^2/r_+^{2n-5}}{4\pi l^2r_+(r_+^2+2k\tilde{\alpha})},\nonumber
\end{eqnarray}
while in the limit of $q \rightarrow 0$, $T_H$ reduces to uncharged case.

Another important thermodynamic quantity is the entropy of black hole. In general
relativity, the entropy of black hole satisfy the so-called area formula, namely
entropy equals to one quarter of horizon area \cite{Hawking:1974rv}. However, the
area law of entropy is not satisfied in general in higher derivative
gravity \cite{Jacobson:1993xs}. Using the standard formula for
entropy $S=\int T_H^{-1}(\frac{\partial M}{\partial r_+})_Q dr_+$, we get
\begin{eqnarray}
S=\frac{\Sigma_k}{4} r_+^{n-1}[1+\frac{(n-1)}{(n-3)}\frac{2\tilde{\alpha}k}{r_+^2}],\label{eq:21a}
\end{eqnarray}
where
\begin{eqnarray}
(\frac{\partial M}{\partial r_+})_Q&=&\frac{(n-1)\Sigma_{k}}{16\pi}r_+^{n-5}[\frac{nr_+^4}{l^2}+(n-2)kr_+^2
+(n-4)k^2\tilde{\alpha}+\frac{4\beta^2r_+^4}{n-1}\nonumber\\
&-&\frac{2\sqrt{2}\beta }{(n-1)r_+^{n-5}}\sqrt{2\beta^2r_+^{2n-2}+(n-1)(n-2)q^2}]\nonumber\\
&=&\frac{(n-1)\Sigma_{k}}{4} r_+^{n-4}(r_+^2+2k\tilde{\alpha})T_H.\label{eq:22a}
\end{eqnarray}

In Eq.~(\ref{eq:19a}) $r_+$ is the real root of Eq.~(\ref{eq:21a}) which is a function of $S$.
One may regard the parameter $S$ and $Q$ as a complete set extensive parameters for the mass $M(S,Q)$
and define the intensive parameters conjugate to them \cite{Dehghani:2008qr}.
These quantities are temperature and the electric potential
\begin{eqnarray}
T_H=(\frac{\partial M}{\partial S})_Q, \quad \Phi=(\frac{\partial M}{\partial Q})_S.\label{eq:23a}
\end{eqnarray}
One can find that the intensive quantities $T_H$ and $\Phi$ in Eq.~(\ref{eq:23a}) is consistent with
Eq.~(\ref{eq:16a}) and Eq.~(\ref{eq:20a}), respectively. Hence, the relevant thermodynamic
quantities Eq.~(\ref{eq:16a}) and Eq.~(\ref{eq:20a}) satisfy the first law of thermodynamics
\begin{eqnarray}
dM=T_HdS+\Phi dQ\label{eq:24a}
\end{eqnarray}

The local stability of black hole is determined by the sign of its heat capacity. If the heat
capacity is positive, the black hole is locally stable to thermal fluctuations.
Otherwise the black hole is locally unstable. The heat capacity for a fixed-charge is expressed as
\begin{eqnarray}
C_Q=(\frac{\partial M}{\partial T_H})_Q
=(\frac{\partial M}{\partial r_+})_Q/(\frac{\partial T_H}{\partial r_+})_Q,\label{eq:25a}
\end{eqnarray}
where
\begin{eqnarray}
(\frac{\partial T_H}{\partial r_+})_Q&=&\frac{1}{4\pi r_+^2(r_+^2+2k\tilde{\alpha})^2}
[\frac{nr_+^6}{l^2}+\frac{6n\tilde{\alpha}kr_+^4}{l^2}-(n-2)kr_+^4
-(n-8)k^2\tilde{\alpha}r_+^2\nonumber\\
&-&2(n-4)k^3\tilde{\alpha}^2]+\frac{1}{(n-1)\pi(r_+^2+2k\tilde{\alpha})^2}\left\{r_+^2(r_+^2
+6k\tilde{\alpha})\beta^2\right.\nonumber\\
&+&\left.\frac{\sqrt{2}\beta J(r)}{2r_+^{n-1}\sqrt{2\beta^2r_+^{2n-2}+(n-1)(n-2)q^2}}\right\},\label{eq:26a}
\end{eqnarray}
where $J(r)=(n-1)(n-2)q^2r_+^2((n-2)r_+^2+2k(n-4)\tilde{\alpha})-2r_+^{2n}(r_+^2
+6k\tilde{\alpha})\beta^2$.

\section{Stability of five dimensional black holes}
\label{33s}

Note that the expression for heat capacity $C_Q$ looks like very complicated. In order to
perform the stability further, in this section, we explore the Gauss-Bonnet-Born-Infeld black
holes in five dimension spacetimes. Then, the black hole solution Eq.~(\ref{eq:13a}) becomes
\begin{eqnarray}
f(r)&=&k+\frac{r^2}{2\tilde{\alpha}}\left\{1-\Big[1-\frac{4\tilde{\alpha}}{l^2}+\frac{4\tilde{\alpha} m}{r^4}
-\frac{4\tilde{\alpha}\beta^2}{3}(1-\sqrt{1+\frac{3q^2}{\beta^2r^6}})\right.\nonumber\\
&-&\left.\frac{6\tilde{\alpha}q^2}{r^6}
\times _2F_1[\frac{1}{3},\frac{1}{2},\frac{4}{3},-\frac{3q^2}{\beta^2 r^{6}}]\Big]^{1/2}\right\}.\label{eq:27a}
\end{eqnarray}
Using the Eq.~(\ref{eq:22a}) and Eq.~(\ref{eq:26a}), therefore, the heat capacity $C_Q$ is obtained
\begin{eqnarray}
C_Q&=&\frac{3\Sigma _k r_+^3(r_+^2+2k\tilde{\alpha})^2\sqrt{1+\frac{3q^2}{\beta^2 r_+^6}}}{\Upsilon+\Xi}\nonumber\\
&\times&[6r_+^2+l^2(3k+2\beta^2r_+^2(1-\sqrt{1+\frac{3q^2}{\beta^2 r_+^6}}))],\label{eq:28a}
\end{eqnarray}
where $\Upsilon=8r_+^4[(3+l^2\beta^2)\sqrt{1+\frac{3q^2}{\beta^2 r_+^6}}-l^2\beta^2](r_+^2+6k\tilde{\alpha})$ and
$\Xi=12l^2[kr_+^2(2k\tilde{\alpha}-r_+^2)\sqrt{1+\frac{3q^2}{\beta^2 r_+^6}}+4q^2]$.
One can see that these thermodynamic properties drastically depend on the event horizon structure $k$.
According to the classification of event horizon structures $k=0$ and $k\pm1$, below each case
will be discussed respectively.

\subsection{The case of $k=0$}
\indent

For $k=0$ and $n=4$, we have
\begin{eqnarray}
T_H&=&\frac{1}{3\pi l^2 r_+^2}[(3+l^2\beta^2)r_+^3-\beta l^2\sqrt{\beta^2r_+^6+3q^2}],\quad
S=\frac{\Sigma_k r_+^3}{4},\label{eq:28b}\\
C_Q&=&\frac{3\Sigma _k r_+^9[3+l^2\beta^2(1-\sqrt{1+\frac{3q^2}{\beta^2 r_+^6}})]
\sqrt{1+\frac{3q^2}{\beta^2 r_+^6}}}{4r_+^6[(3+l^2\beta^2)\sqrt{1+\frac{3q^2}{\beta^2 r_+^6}}-l^2\beta^2]+24l^2q^2}\nonumber\\
&=&\frac{3\Sigma _k r_+^6\sqrt{1+\frac{3q^2}{\beta^2 r_+^6}}}{4r_+^6[(3+l^2\beta^2)\sqrt{1+\frac{3q^2}{\beta^2 r_+^6}}
-l^2\beta^2]+24l^2q^2}T_H.\label{eq:29a}
\end{eqnarray}

It is interesting to notice that these thermodynamic properties are independent of the coefficient
$\tilde{\alpha}$ and the entropy of black hole is proportional to the area of the horizon. From
the Eq.~(\ref{eq:28b}), the radius of the extremal black hole $T_H=0$ is obtained
$r_0=\sqrt[6]{\frac{q^2\beta^2 l^4}{3+2l^2\beta^2}}$ which is the only root of Eq.~(\ref{eq:28b}). In Fig.~$1$,
we draw the temperatures $T_H$ of black holes with different values of the parameter $\beta=2$, 1 and 0.1.
For instance, the temperature $T_H$ vanishes at $r_+=r_0 \approx 0.76$ for $l=\beta=1$. Then it goes
to positive infinity as $r_+ \rightarrow \infty$. Furthermore, this similar trait of the temperatures $T_H$
of black holes is universal for different values of parameter $l$.
On the other hand, the heat capacity $C_Q$ disappears for $T_H=0$ which is
presented in Eq.~(\ref{eq:29a}). One can see that $C_Q$ is positive provided $r_+>r_0$ (see Fig. 1).
We therefore conclude that the black holes are stable against fluctuations as
they are shown in \cite{Dehghani:2006ke}.

\begin{figure}[htb]
\centering
\subfigure[$\beta=1$ and $l=1$]{
\label{fig:subfig:a} 
\includegraphics{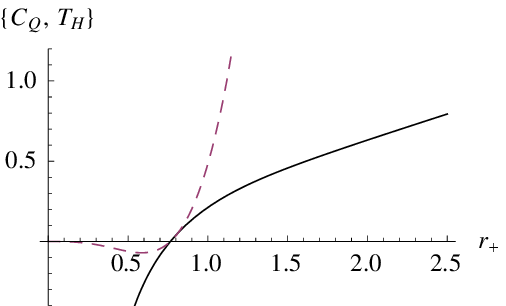}}%
\hfill%
\subfigure[$\beta=0.1$ and $l=2$]{
\label{fig:subfig:b} 
\includegraphics{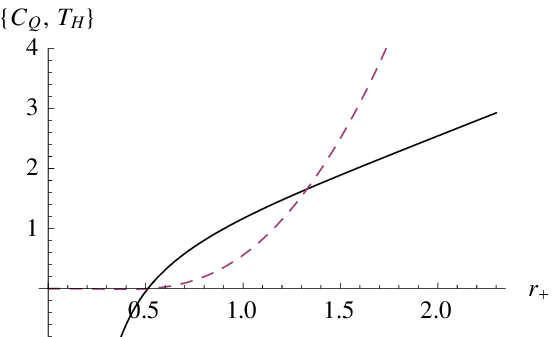}}%
\hfill%
\subfigure[$\beta=1$ and $l=0.5$]{
\label{fig:subfig:c} 
\includegraphics{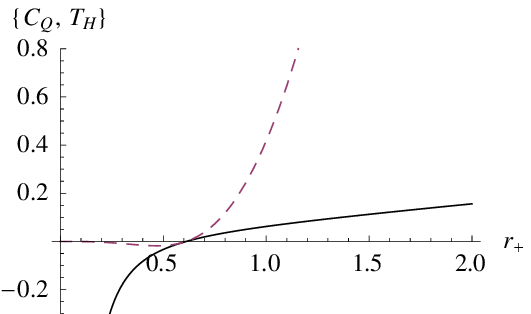}}
\caption{ Temperature $T_H$ (solid curves) and heat capacity $C_Q$ (dashed curves) versus horizon
radius $r_+$ for $n=4$, $q=1$ and $k=0$. Here we choose $\Sigma_k=1$.}
\label{fig:subfig} 
\end{figure}

Considering black holes in flat space, we obtain the temperature $T_H$ of black holes
in case of $l \rightarrow \infty$
\begin{eqnarray}
T_H=\frac{\beta^2r_+}{3\pi}(1-\sqrt{1+\frac{3q^2}{\beta^2 r_+^6}}),\label{eq:29b}
\end{eqnarray}
Apparently the temperature $T_H$ is always negative in the whole range of $r_+$.
Henceforth there don't exist black holes in flat space for $k=0$.

\subsection{The case of $k=1$}
\indent
In this subsection we explore some physical aspects of the black holes with
positive constant curvature hypersurface horizon. For $n=4$, we find that this class of
solutions was also studied in \cite{Aiello:2004rz, Wiltshire:1988uq}. While, we are concerned
with the thermodynamic properties of black holes which have been never discussed before.
One can see that there maybe exist extremal black holes for $T_H=0$ in Eq.~(\ref{eq:20a}),
which is expressed as
\begin{eqnarray}
T_H=\frac{1}{6\pi l^2(r_{ext}^2+2\tilde{\alpha})}[6r_{ext}^3+3l^2r_{ext}+2l^2\beta^2 r_{ext}^3
(1-\sqrt{1+\frac{3q^2}{\beta^2r_{ext}^6}})]=0.\label{eq:30a}
\end{eqnarray}
For simplicity, here we set the parameters $l=1$ and $q=1$. Then, the Eq.~(\ref{eq:30a})
can be rewritten as
\begin{eqnarray}
4(3+2\beta^2)R_{ext}^3+4(3+\beta^2)R_{ext}^2+3R_{ext}-4\beta^2=0.\label{eq:31a}
\end{eqnarray}
where $r_{ext}^2 = R_{ext}$, namely $R_{ext}\geq 0$. Then, the discriminant of this
cubic equation is given by
\begin{eqnarray}
\Delta=48\beta^2(216+4095\beta^2+5040\beta^4+1664\beta^6).\label{eq:32a}
\end{eqnarray}
Clearly, it is always positive for parameter $\beta$. Therefore, there only exist one positive
real root of this cubic equation. The radius of the event horizon for extremal black hole is given by
\begin{eqnarray}
r_{ext}^2&=&\frac{1}{6(3+2\beta^2)}\left\{-2(3+\beta^2)+[J(r)-3\beta(3+2\beta^2)\sqrt{\Gamma(\beta)}]^{1/3}\right.\nonumber\\
&+&\left.[J(\beta)+3\beta(3+2\beta^2)\sqrt{\Gamma(\beta)}]^{1/3}\right\},\nonumber\\\label{eq:33a}
\end{eqnarray}
where $\Gamma(\beta)=3(216+4095\beta^2+5040\beta^4+1664\beta^6)$ and $J(\beta)=27+999\beta^2
+1278\beta^4+424\beta^6$.
In Fig.~2, we plot the temperature $T_H$ with different values of parameters $\tilde{\alpha}$ and $\beta$.
Obviously, the temperature $T_H$ vanishes at $r_+=r_{ext}$, and then goes to positive infinity
as radius increases. This trait does not change even though the black hole solutions take various
values for parameters $\tilde{\alpha}$ and $\beta$.

Based on Eq.~(\ref{eq:19a}), therefore the mass of extremal black hole is expressed in terms
of $r_{ext}$ as
\begin{eqnarray}
M_{ext}&=&\frac{3\Sigma_k r_{ext}^2}{16\pi}\left\{1+\frac{\tilde{\alpha}}{r_{ext}^2}
+r_{ext}^2+\frac{\beta^2r_{ext}^2}{3}-\frac{\beta}{3r_{ext}}
\sqrt{\beta^2r_{ext}^6+3}\right.\nonumber\\
&+&\left.\frac{3}{2r_{ext}^4}\times_2F_1[\frac{1}{3},\frac{1}{2},\frac{4}{3},
-\frac{3}{\beta^2r_{ext}^6}]\right\}.\label{eq:34a}
\end{eqnarray}
Note that if $m>m_{ext}$, there are more than one horizon while there will be
degenerate horizon at $r=r_{ext}$ for $m=m_{ext}$. But for $m<m_{ext}$, no horizon
exists and we are left with a naked singularity.

With regard to the stability of black holes, $C_Q$ has been demonstrated
by $(\frac{\partial M}{\partial r_+})_Q/(\frac{\partial T_H}{\partial r_+})_Q$ in Eq.~(\ref{eq:25a}).
Since the function $(\frac{\partial M}{\partial r_+})_Q$
is always non-negative for $T_H \geq 0$, the sign of heat capacity $C_Q$ is determined by the one of
function $(\frac{\partial T_H}{\partial r_+})_Q$. From Fig.~2, one see that $T_H$
vanishes at $r_+=r_{ext}$, and then goes to positive infinity as radius $r_+$ increases.
Therefore, $C_Q$ is always positive when $r_+>r_{ext}$. In this case, the black holes are
locally stable in the region $r_+>r_{ext}$.

\begin{figure}[htb]
\centering
\subfigure[$\tilde{\alpha}=1$ and $\beta=1$]{
\label{fig:subfig:a} 
\includegraphics{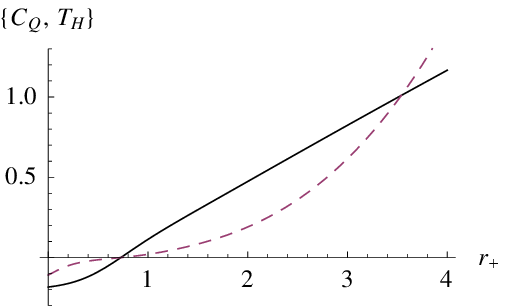}}%
\hfill%
\subfigure[$\tilde{\alpha}=1$ and $\beta=0.5$]{
\label{fig:subfig:b} 
\includegraphics{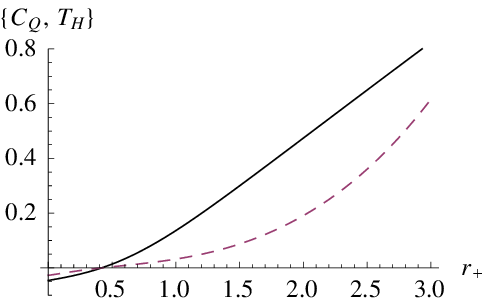}}%
\hfill%
\subfigure[$\tilde{\alpha}=0.1$ and $\beta=1$]{
\label{fig:subfig:c} 
\includegraphics{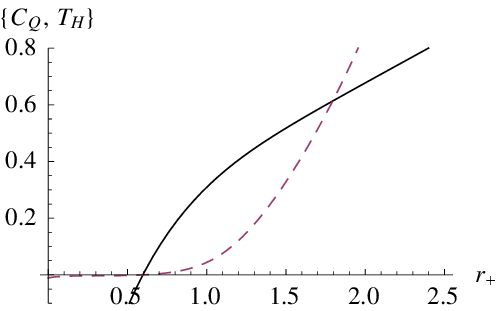}}
\caption{ Temperature $T_H$ (solid curves) and heat capacity $C_Q$ (dashed curves) versus horizon
radius $r_+$ for $l=1$, $n=4$, $q=1$ and $k=1$. Here we choose $\Sigma_k=0.01$.}
\label{fig:subfig} 
\end{figure}

In case of $\Lambda \rightarrow 0$, namely $l^2 \rightarrow \infty$, the Gauss-Bonnet-Born-Infeld
black holes reduces to the flat case. And then, a asymptotically flat black hole solution
can be easily obtained by taking $k=1$. Hence, the temperature $\bar{T}_H$ of this new solution is given by
\begin{eqnarray}
\bar{T}_H=\frac{r_+}{6\pi(r_+^2+2\tilde{\alpha})}[3+2\beta^2 r_+^2
(1-\sqrt{1+\frac{3q^2}{\beta^2r_+^6}})].\label{eq:34b}
\end{eqnarray}
The extremal radius $\bar{r}_{ext}$ can be straightforwardly written as
$\bar{r}_{ext}=\frac{(\sqrt{9+64\beta^4}-3)^{1/2}}{2\sqrt{2}\beta}$.
According to Eq.~(\ref{eq:19a}), the mass of black hole in case of $l \rightarrow \infty$ is given by
\begin{eqnarray}
\bar{M}_{ext}&=&\frac{3\Sigma_k \bar{r}_{ext}^2}{16\pi}\left\{1+\frac{\tilde{\alpha}}{\bar{r}_{ext}^2}
+\frac{\beta^2\bar{r}_{ext}^2}{3}-\frac{\beta}{3\bar{r}_{ext}}
\sqrt{\beta^2\bar{r}_{ext}^6+3}\right.\nonumber\\
&+&\left.\frac{3}{2\bar{r}_{ext}^4}\times_2F_1[\frac{1}{3},\frac{1}{2},\frac{4}{3},
-\frac{3}{\beta^2\bar{r}_{ext}^6}]\right\}.\label{eq:34a}
\end{eqnarray}

Different from the AdS case, the temperature $\bar{T}_H$ starts from zero at $r_+=\bar{r}_{ext}$ increases
sharply reaches a local maximum at $r_+=r_m$, and then decreases gradually to zero as $r_+ \rightarrow \infty$.
The temperature $\bar{T}_H$ with different values of parameter $\beta$ and
coefficient $\tilde{\alpha}$ is plotted in Fig.~3.
Based on Eq.~(\ref{eq:25a}) $C_Q=(\frac{\partial M}{\partial r_+})_Q/(\frac{\partial T_H}{\partial r_+})_Q$,
the sign of heat capacity $C_Q$ is only determined by $(\frac{\partial \bar{T}_H}{\partial r_+})_Q$ and $C_Q$
also disappears when $\bar{T}_H$ is equal to zero. Therefore, the $C_Q$ maintains positive in the
region $r_{ext}< r_+ < r_m$, and then becomes negative in the region $r_+ > r_m$. As a result, the black holes
are locally stable in the domain $r_{ext}< r_+ < r_m$ and locally unstable provided $r_+ > r_m$.
Obviously, the cosmological constant $\Lambda$ plays an important role in the distribution of stable
regions of Gauss-Bonnet-Born-Infeld black holes.

\begin{figure}[htb]
\centering
\subfigure[$\alpha=1$ and $\beta=1$]{
\label{fig:subfig:b} 
\includegraphics{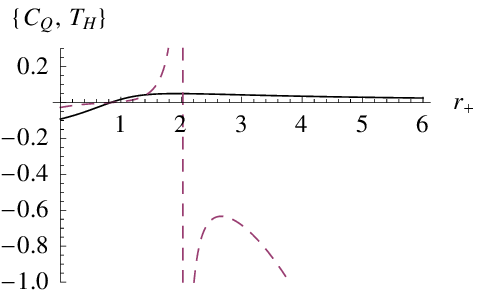}}%
\hfill%
\subfigure[$\alpha=1$ and $\beta=0.5$]{
\label{fig:subfig:b} 
\includegraphics{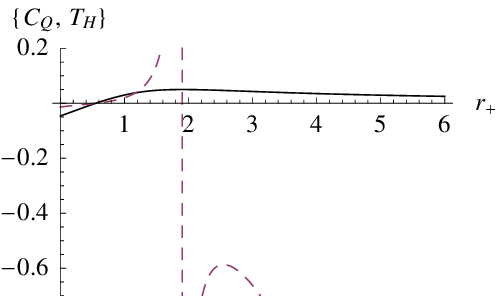}}%
\hfill%
\subfigure[$\alpha=0.1$ and $\beta=1$]{
\label{fig:subfig:c} 
\includegraphics{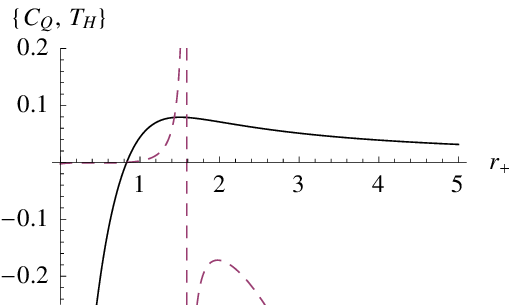}}
\caption{ Temperature $T_H$ (solid curves) and heat capacity $C_Q$ (dashed curves) versus horizon
radius $r_+$ for $n=4$, $q=1$ and $k=1$. Here we choose $\Sigma_k=1/800$.}
\label{fig:subfig} 
\end{figure}

\subsection{The case of $k=-1$}
\indent
Now, we turn to the case the horizon is a negative constant curvature hypersurface.
In five dimensional spacetimes, the temperature of black holes Eq.~(\ref{eq:20a}) becomes
\begin{eqnarray}
T_H=\frac{1}{6\pi l^2(r_{+}^2-2\tilde{\alpha})}[6r_{+}^3-3l^2r_{+}+2l^2\beta^2 r_{+}^3
(1-\sqrt{1+\frac{3q^2}{\beta^2r_{+}^6}})].\label{eq:35a}
\end{eqnarray}
Then, the existence of extremal black holes depends on the existence of positive root(s)
for $T_H=0$, which reduces to
\begin{eqnarray}
6r_{ext}^3-3l^2r_{ext}+2l^2\beta^2 r_{ext}^3-2l^2\beta\sqrt{\beta^2r_{ext}^6+3q^2}=0.\label{eq:36a}
\end{eqnarray}

Adopted the same approach above, we order the parameters $l=1$ and $q=1$. Then,
the Eq.~(\ref{eq:36a}) can be rewritten as
\begin{eqnarray}
4(3+2\beta^2)R_{ext}^3-4(3+\beta^2)R_{ext}^2+3R_{ext}-4\beta^2=0,\label{eq:37a}
\end{eqnarray}
where $R_{ext}$ is defined as $r_{ext}^2$, namely, $R_{ext} \geq 0$. Then, the discriminant of
this cubic equation is
\begin{eqnarray}
\Delta_*=48\beta^2(-216+3663\beta^2+5328\beta^4+1792\beta^6).\label{eq:38a}
\end{eqnarray}
Different from the case $k=1$, the discriminant $\Delta_*$ can be take negative or positive values for
different values of parameter $\beta$. However, the root of Eq.~(\ref{eq:38a}) is not real and
is located between 0.2335 and 0.2336. Therefore, we discuss the two case respectively.

If the parameter $\beta \geq 0.2336$, the discriminant $\Delta_*$ is positive. Then, the only one
real root is obtained
\begin{eqnarray}
r_{ext}^2&=&\frac{1}{12(3+2\beta^2)}\left\{4(3+\beta^2)+[J(\beta)-6(3+2\beta^2)\sqrt{\Delta_*}]^{1/3}\right.\nonumber\\
&+&\left.[J(\beta)+6(3+2\beta^2)\sqrt{\Delta_*}]^{1/3}\right\},\label{eq:39a}
\end{eqnarray}
where $J(\beta)=8(-27+945\beta^2+1314\beta^4+440\beta^6)$. In case of $\beta \leq 0.2335$, namely, $\Delta_*<0$,
there exist three different real roots. They are give by
\begin{eqnarray}
r_{e1}^2&=&\frac{1}{3(3+2\beta^2)}[3+\beta^2-\sqrt{9+6\beta^2+4\beta^4}\cos(\theta/3)],\nonumber\\
r_{e2}^2&=&\frac{1}{3(3+2\beta^2)}\left\{3+\beta^2+\frac{1}{2}\sqrt{9+6\beta^2
+4\beta^4}[\cos(\theta/3)
-\sqrt{3}\sin(\theta/3)]\right\},\nonumber\\
r_{e3}^2&=&\frac{1}{3(3+2\beta^2)}\left\{3+\beta^2+\frac{1}{2}\sqrt{9+6\beta^2
+4\beta^4}[\cos(\theta/3)
+\sqrt{3}\sin(\theta/3)]\right\},\label{eq:40a}
\end{eqnarray}
where $\theta=\arccos U$, $U=-\frac{-27+945\beta^2+1314\beta^4+440\beta^6}{(9+6\beta^2+4\beta^4)^{3/2}}$.
It is worth to mention that the quantities $r_{e1}^2$, $r_{e2}^2$ and $r_{e3}^2$ is only the three roots of
Eq.~(\ref{eq:37a}). But not all roots satisfy the equation of temperature $T_H$ Eq.~(\ref{eq:35a}).
We find that only $r_{e3}^2$ is the root of Eq.~(\ref{eq:35a}).
In addition, the horizon radius must obey
\begin{eqnarray}
r_+^2 \geq 2\tilde{\alpha}.\label{eq:41a}
\end{eqnarray}
Therefore, the radius of extremal black hole has a constraint $r_{ext}^2 \geq 2\tilde{\alpha}$.
The graph of $T_H$ is plotted in Fig.~4. The temperature $T_H$ blows up
at $r_+=\sqrt{2\tilde{\alpha}}$ and changes sign at $r_+=r_{ext}$ ($\Delta_*>0$)
or $r_+=r_{e3}$ ($\Delta_*<0$). Then, it gradually increases as $r_+ \rightarrow \infty$.

Now, let us discuss the stability of black holes. In Fig.~4, we have plotted $C_Q$ as a function
of horizon radius $r_+$. We see from the figure that the heat capacity $C_Q$ is positive if $r_+$
is larger than the radius of extremal black hole $r_{ext}$. It means that
the black holes are locally stable in the region $r_+>r_{ext}$.

\begin{figure}[htb]
\centering
\subfigure[$\tilde{\alpha}=0.1$ and $\Delta_*>0(\beta=1)$]{
\label{fig:subfig:a} 
\includegraphics{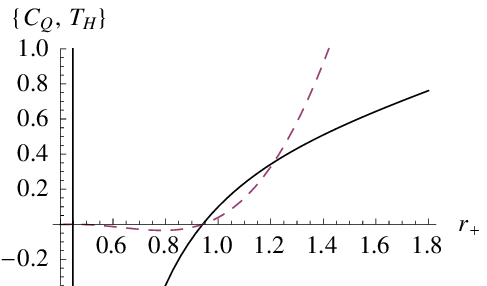}}%
\hfill%
\subfigure[$\tilde{\alpha}=0.1$ and $\Delta_*<0(\beta=0.1)$]{
\label{fig:subfig:b} 
\includegraphics{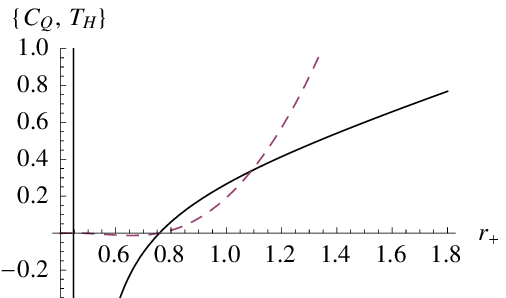}}
\caption{ Temperature $T_H$ (solid curves) and heat capacity $C_Q$ (dashed curves) versus horizon
radius $r_+$ for $n=4$, $q=1$, $l=1$ and $k=-1$. Here we choose $\Sigma_k=1$.}
\label{fig:subfig} 
\end{figure}

However, the case of $\Lambda=0$ is quite different from corresponding one above.
Here we also let $q=1$. Then, the temperature $T_H$ is
\begin{eqnarray}
T_H=\frac{1}{6\pi (r_{+}^2-2\tilde{\alpha})}[-3r_{+}+2\beta^2 r_{+}^3
(1-\sqrt{1+\frac{3}{\beta^2r_{+}^6}})].\label{eq:42a}
\end{eqnarray}

\begin{figure}
\includegraphics{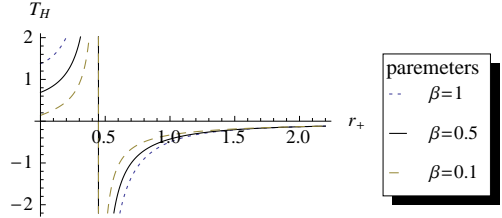}
\caption{ Temperature $T_H$ versus $r_+$ for $n=4$, $q=1$ $\alpha=0.1$ and $k=-1$.}
\end{figure}

Since the horizon radius also satisfies $r_+^2 \geq 2\tilde{\alpha}$, we find that this
equation Eq.~(\ref{eq:42a}) does not have positive root. Hence, we conclude that for $k=-1$,
the asymptotically flat black holes are unstable in the whole range of $r_+$.

\section{concluding remarks}
\label{44s}

In this paper, we have constructed static topological black hole solutions of Gauss-Bonnet-Born-Infeld
action in the presence of cosmological constant. Then, we discussed the thermodynamic properties
of black holes including gravitational mass, Hawking temperature and entropy of black holes.
We also notice that the entropy of Gauss-Bonnet-Born-Infeld black holes get no correction from the
Born-Infeld gauge field.

Later, we performed the stability analysis of these topological black holes in five dimensional
spacetimes for convenience. For $k=0$, all the thermodynamic quantities of the black
holes don't depend on the Gauss-Bonnet coefficient $\tilde{\alpha}$ and are the same as those of
black holes in Einstein-Born-Infeld gravity although the two black hole solutions are quit different.
For the horizon is negative constant hypersurface, the cosmological constant $\Lambda$ plays a important
role in the distribution of regions for the stability of black holes.
the Gauss-Bonnet-Born-Infeld black holes in AdS space are thermodynamically stable provided $r_+>r_{ext}$,
while corresponding ones in flat space are unstable in the whole range of $r_+$.
For the horizon is positive constant hypersurface, unlike the uncharged Gauss-Bonnet black holes
that there don't exist extremal black holes when $k=1$ \cite{Cai:2001dz},
the extremal black holes are also existence in Gauss-Bonnet-Born-Infeld gravity.
In addition, the asymptotically flat Gauss-Bonnet-Born-Infeld black holes are thermodynamically
stable in the region $r_+>r_{ext}$.  However, since the existence of cosmological constant $\Lambda$,
the black holes becomes unstable beyond the location $r_+=r_m$ where the temperature reaches local maximum.

{\bf Acknowledgment }
This work  has been supported by the Natural Science Foundation of China
under grant No.10875060.


\begin{thebibliography}{}

\bibitem{Fradkin:1985qd}
  E.~S.~Fradkin and A.~A.~Tseytlin,
  Phys.\ Lett.\  B {\bf 163}, 123 (1985);
\bibitem{Bergshoeff:1987at}
  E.~Bergshoeff, E.~Sezgin, C.~N.~Pope and P.~K.~Townsend,
  Phys.\ Lett.\  {\bf 188B}, 70 (1987).
\bibitem{Metsaev:1987qp}
  R.~R.~Metsaev, M.~Rakhmanov and A.~A.~Tseytlin,
  Phys.\ Lett.\  B {\bf 193}, 207 (1987).
\bibitem{Lust:1989tj}
  D.~Lust and S.~Theisen,
  Lect.\ Notes Phys.\  {\bf 346}, 1 (1989).
\bibitem{Myers:1987yn}
  R.~C.~Myers,
  Phys.\ Rev.\  D {\bf 36}, 392 (1987).
\bibitem{Alekseev:1997wy}
  S.~O.~Alekseev and M.~V.~Pomazanov,
  Grav.\ Cosmol.\  {\bf 3}, 191 (1997).
\bibitem{Maeda:2009uy}
  K.~i.~Maeda, N.~Ohta and Y.~Sasagawa,
  Phys.\ Rev.\  D {\bf 80}, 104032 (2009)
  [arXiv:0908.4151 [hep-th]].
\bibitem{Dehghani:2009zzb}
  M.~H.~Dehghani and R.~Pourhasan,
  Phys.\ Rev.\  D {\bf 79}, 064015 (2009)
  [arXiv:0903.4260 [gr-qc]].
\bibitem{Lovelock:1971yv}
  D.~Lovelock,
  J.\ Math.\ Phys.\  {\bf 12}, 498 (1971).
\bibitem{Zwiebach:1985uq}
  B.~Zwiebach,
  Phys.\ Lett.\  B {\bf 156}, 315 (1985);
  B.~Zumino,
  Phys.\ Rept.\  {\bf 137}, 109 (1986).
\bibitem{Wheeler:1985qd}
  J.~T.~Wheeler,
  Nucl.\ Phys.\  B {\bf 273}, 732 (1986);
  J.~T.~Wheeler,
  Nucl.\ Phys.\  B {\bf 268}, 737 (1986).
\bibitem{Boulware:1985wk}
  D.~G.~Boulware and S.~Deser,
  Phys.\ Rev.\ Lett.\  {\bf 55}, 2656 (1985).
\bibitem{Wiltshire:1985us}
  D.~L.~Wiltshire,
  Phys.\ Lett.\  B {\bf 169}, 36 (1986).
\bibitem{Myers:1988ze}
R.~C.~Myers and J.~Z.~Simon,
  Phys.\ Rev.\  D {\bf 38}, 2434 (1988);
  I.~P.~Neupane,
  Phys.\ Rev.\  D {\bf 69}, 084011 (2004)
  [arXiv:hep-th/0302132].
\bibitem{Cai:2001dz}
  R.~G.~Cai,
  Phys.\ Rev.\  D {\bf 65}, 084014 (2002)
  [arXiv:hep-th/0109133];
  Y.~M.~Cho and I.~P.~Neupane,
  Phys.\ Rev.\  D {\bf 66}, 024044 (2002)
  [arXiv:hep-th/0202140].
 \bibitem{Cvetic:2001bk}
  M.~Cvetic, S.~Nojiri and S.~D.~Odintsov,
  Nucl.\ Phys.\  B {\bf 628}, 295 (2002)
  [arXiv:hep-th/0112045];
  J.~E.~Lidsey, S.~Nojiri and S.~D.~Odintsov,
  JHEP {\bf 0206}, 026 (2002)
  [arXiv:hep-th/0202198];
  S.~Nojiri and S.~D.~Odintsov,
  Phys.\ Lett.\  B {\bf 521}, 87 (2001)
  [Erratum-ibid.\  B {\bf 542}, 301 (2002)]
  [arXiv:hep-th/0109122].
\bibitem{Born:1934gh}
  M.~Born and L.~Infeld,
  Proc.\ Roy.\ Soc.\ Lond.\  A {\bf 144}, 425 (1934).
\bibitem{Hoffmann:1935ty}
  B.~Hoffmann,
  Phys.\ Rev.\  {\bf 47} (1935) 877.
\bibitem{Aiello:2004rz}
  M.~Aiello, R.~Ferraro and G.~Giribet,
  Phys.\ Rev.\  D {\bf 70}, 104014 (2004)
  [arXiv:gr-qc/0408078].
\bibitem{Garica:1984}
  A.~Garcia, H.~Salazar, and  J.~F.~Plebanski,
   Nuovo.\ Cimento.\ Soc.\ Ital.\ Fis.\ A {bf 84}, 65 (1984)
\bibitem{Cai:2004eh}
  R.~G.~Cai, D.~W.~Pang and A.~Wang,
  Phys.\ Rev.\  D {\bf 70}, 124034 (2004)
  [arXiv:hep-th/0410158];
  S.~Fernando and D.~Krug,
  Gen.\ Rel.\ Grav.\  {\bf 35}, 129 (2003)
  [arXiv:hep-th/0306120];
  T.~K.~Dey,
  Phys.\ Lett.\  B {\bf 595}, 484 (2004)
  [arXiv:hep-th/0406169].
\bibitem{Sheykhi:2008rt}
  A.~Sheykhi,
  Int.\ J.\ Mod.\ Phys.\  D {\bf 18}, 25 (2009)
  [arXiv:0801.4112 [hep-th]];
  A.~Sheykhi,
  Phys.\ Lett.\  B {\bf 662}, 7 (2008)
  [arXiv:0710.3827 [hep-th]];
  A.~Sheykhi and N.~Riazi,
  Phys.\ Rev.\  D {\bf 75}, 024021 (2007)
  [arXiv:hep-th/0610085].
\bibitem{Myung:2008eb}
  Y.~S.~Myung, Y.~W.~Kim and Y.~J.~Park,
  Phys.\ Rev.\  D {\bf 78}, 084002 (2008)
  [arXiv:0805.0187 [gr-qc]];
  S.~Fernando,
  Phys.\ Rev.\  D {\bf 74}, 104032 (2006)
  [arXiv:hep-th/0608040].
\bibitem{Wiltshire:1988uq}
  D.~L.~Wiltshire,
  Phys.\ Rev.\  D {\bf 38}, 2445 (1988).
\bibitem{Dehghani:2008qr}
  M.~H.~Dehghani, N.~Alinejadi and S.~H.~Hendi,
  Phys.\ Rev.\  D {\bf 77}, 104025 (2008)
  [arXiv:0802.2637 [hep-th]].
\bibitem{Dehghani:2006zi}
  M.~H.~Dehghani, S.~H.~Hendi, A.~Sheykhi and H.~Rastegar Sedehi,
  JCAP {\bf 0702}, 020 (2007)
  [arXiv:hep-th/0611288].
\bibitem{Dehghani:2006ke}
  M.~H.~Dehghani and S.~H.~Hendi,
  Int.\ J.\ Mod.\ Phys.\  D {\bf 16}, 1829 (2007)
  [arXiv:hep-th/0611087].
  M.~H.~Dehghani and S.~H.~Hendi,
  Gen.\ Rel.\ Grav.\  {\bf 41}, 1853 (2009)
  [arXiv:0903.4259 [hep-th]].
\bibitem{Hawking:1974rv}
  S.~W.~Hawking,
  Nature {\bf 248}, 30 (1974).
  J.~D.~Bekenstein,
  Phys.\ Rev.\  D {\bf 7}, 2333 (1973);
  G.~W.~Gibbons and S.~W.~Hawking,
  Phys.\ Rev.\  D {\bf 15}, 2738 (1977).
\bibitem{Jacobson:1993xs}
  T.~Jacobson and R.~C.~Myers,
  Phys.\ Rev.\ Lett.\  {\bf 70}, 3684 (1993)
  [arXiv:hep-th/9305016].

\end{thebibliography}
\end{document}